\documentclass[letterpaper,10pt]{article} 

\usepackage{opticameet3} 

\newcommand\authormark[1]{\textsuperscript{#1}}

\usepackage{amsmath,amssymb}
\usepackage[colorlinks=true,bookmarks=false,citecolor=blue,urlcolor=blue]{hyperref} 

\newcommand{\secInLine}[1]{\textbf{#1}.}

\usepackage{acro} 
\DeclareAcronym{dc}{
  short=DC,
  long=data centre,
}

\DeclareAcronym{dcn}{
  short=DCN,
  long=data centre network,
}

\DeclareAcronym{jsd}{
  short=JSD,
  long=Jensen-Shannon distance,
}

\usepackage{siunitx} 
\usepackage{numprint} 

\usepackage[font=footnotesize, skip=2pt]{caption}

\usepackage{caption}
\usepackage{algorithm}
\usepackage{algorithmic}

\usepackage{wrapfig} 

\begin{document}

\title{A Vectorised Packing Algorithm for Efficient Generation of Custom Traffic Matrices}


\author{Christoper W. F. Parsonson,\authormark{1,*} Joshua L. Benjamin,\authormark{1} and Georgios Zervas\authormark{1}}

\address{\authormark{1} Optical Networks Group, Electronic \& Electrical Engineering, UCL}

\email{\authormark{*}zciccwf@ucl.ac.uk} 

\begin{abstract}
We propose a new algorithm for generating custom network traffic matrices which achieves $13\times$, $38\times$, and $70\times$ faster generation times than prior work on networks with \num{64}, \num{256}, and \num{1024} nodes respectively.
\end{abstract}

\section{Introduction}

\Acp{dc} have become critical tools for modern computational tasks. 
To meet the ever-increasing demands of \acp{dc}, recent years have seen a growth in the research and development of next-generation \ac{dc} optical systems \cite{khani2021sipml}. However, most researchers rely on simulations, which require the generation of synthetic traffic.
In doing so, they often make overly simplistic assumptions about the characteristics of their generated traffic and develop systems which, in practice, perform poorly under real-world conditions \cite{parsonson2022traffic}. Furthermore, many works omit open-accessing their synthetic traffic or even the methodology used to generate it, bringing problems with reproducibility, benchmarking, and cross-validation. The lack of a reproducible and high-fidelity synthetic traffic generation tool has been a long-standing problem in the \ac{dc} research community.

Prior works \cite{alizadeh2013pfabric, bai2016enabling} have released traffic generators, but these were either intended to be unrealistic, were for specific network topologies, required the cumbersome use of inflexible configuration files, or lacked a reproducibility guarantee. To address this, recent work presented TrafPy; an open source tool for generating reproducible \ac{dc} traffic with custom distributions and characteristics \cite{parsonson2022traffic}. However, the authors only demonstrated traffic generation for \num{64} network nodes; far smaller than the $O($\num{1000}$)$ node \acp{dc} which are becoming increasingly common place. 

In this work, we first show that the original flow source-destination assignment algorithm (`packing', see Section \ref{sec:custom_traffic_matrix_generation}) used in the TrafPy generator is a major bottleneck because its time complexity scales poorly with the number of \ac{dc} nodes $|N|$ for which $|F|$ flows are being generated. This prevents the generation of traffic for large networks. Next, we propose a novel vectorised packing algorithm which fits in with the rest of the authors' traffic generation framework. Finally, we demonstrate our vectorised packer achieving 
$13\times$, $38\times$, and $70\times$ faster generation times than \cite{parsonson2022traffic} on networks with \num{64}, \num{256}, and \num{1024} nodes respectively with up to $\approx5$M traffic flows, with the speed-up factor increasing with the network size.
We expect this work to unlock a new realm of \ac{dc} research at scale and to further facilitate the development of next-generation systems and common platforms for benchmarking networks. We note that while here we focus on generating traffic for optical \acp{dc}, the same traffic generation scheme and vectorised packing algorithm could be re-purposed and applied to any network system.

\section{Custom Traffic Matrix Generation}
\label{sec:custom_traffic_matrix_generation}

\secInLine{Problem statement} 
\Ac{dc} traffic is made up of \textit{flows}. A flow $f$ is fully described by its \textit{size} $f^{s}$ (how much information to send), \textit{arrival time} $f^{a}$ (when the flow requests to be transported through the \ac{dc}, thus giving rise to the \textit{inter-arrival time} in a dynamic multi-flow setting), and \textit{source-destination pair} $f^{p}$ (which machines in the \ac{dc} the flow is requesting to be sent between). In the framework of \cite{parsonson2022traffic}, traffic generation is split into two stages. In the first stage (`shaping and sampling'), custom flow size and inter-arrival time distributions are generated and sampled to attain a set of sizes $\bold{b}^{s}$ and arrival times $\bold{b}^{a}$ for $f \in F$ flows which match the target distributions within some \ac{jsd} threshold\footnote{The \ac{jsd} $\in [0, 1]$ is a measure of how similar two distributions are to one another (lower is more similar), and the \ac{jsd} `threshold' as defined by \cite{parsonson2022traffic} is a constraint on how similar the generated traffic characteristics must be to the target distributions.}. In the second stage (`packing'), given $\bold{b}^{s}$ and $\bold{b}^{a}$, the task is to assign each flow $f \in F$ to a source-destination pair such that some target node distribution (a.k.a. traffic matrix heat map) $\mathbb{P}^{N}$ with nodes $n \in N$ and corresponding source-destination node pairs $p \in P$ is realised as closely as possible without exceeding the load capacity limitations of any node. The authors formulate this task by extracting the fraction of the overall load requested by each pair $p \in P$ into an array, multiplying each element by the overall \ac{dc}'s target load rate to get the per-pair target load rate, and then again multiplying each element by the simulation duration (the time between the first and last flows' arrivals) to get the total amount of information to load onto each pair, $\bold{b}^{p,I}_{target}$, needed in order to achieve the desired target node distribution $\mathbb{P}^{N}$. The packing task is therefore reduced to finding the source-destination node pair assignments for each flow $f \in F$ such that the difference between the actual and the target per-pair total information loads, $\bold{b}^{p,I}_{target} - \bold{b}^{p,I}_{actual}$, is \num{0} \textit{or}, where this is not possible given any incompatibility between the target node distribution $\mathbb{P}^{N}$ and the overall \ac{dc} load rate, to match $\mathbb{P}^{N}$ as closely as possible (see \cite{parsonson2022traffic} for further details).

\begin{figure}[!tp]
    \centering
    \includegraphics[width=0.98\textwidth]{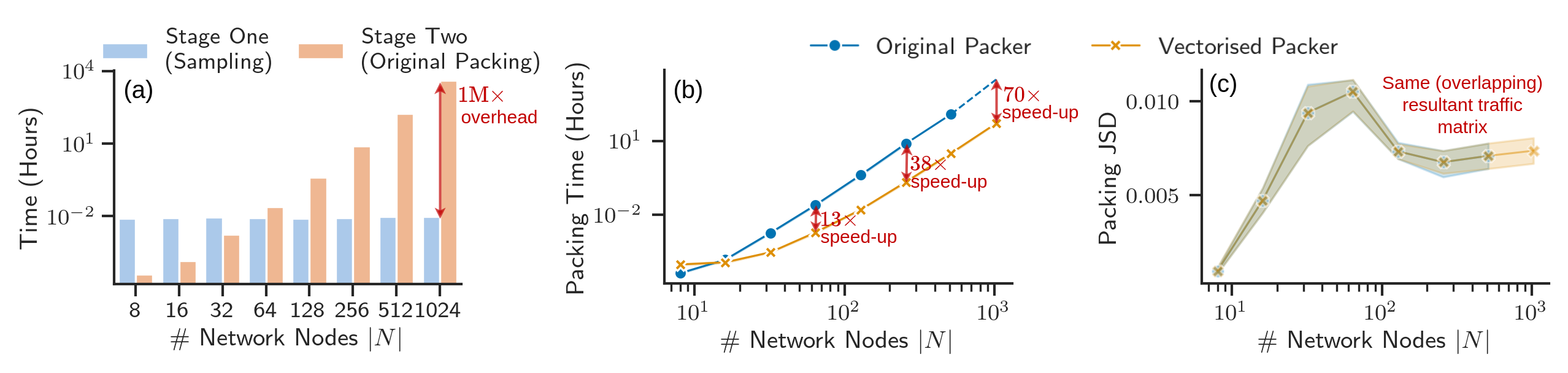}
    \caption{i) (a) The time for stages one (shaping and sampling) and two (packing) when generating flows with the original packing algorithm. ii) The packing (b) time and (c) Jensen-Shannon distance between the target and the generated node distributions for the original and vectorised packing algorithms when generating traffic for networks with different numbers of nodes. (a) shows that the original packing algorithm is the major traffic generation bottleneck of \cite{parsonson2022traffic}. (b) shows that as the number of network nodes is increased, the vectorised packer's speed-up factor over the original algorithm increases. (c) shows that both algorithms achieve the exact same resultant node distribution. Note that the original algorithm's time results for $|N|=1024$ are extrapolations since it would have taken $\approx200$ days to run the packer.}
    \label{fig:packing_time_and_jsd}
\end{figure}

As shown in Figure \ref{fig:packing_time_and_jsd}a, as $|N|$ is increased, stage two (packing) becomes a major bottleneck, taking $\approx$\num{1000000} times longer than stage one for $|N|=$\num{1024}. We therefore focus on optimising stage two.


\secInLine{Original packing algorithm}
The original packing algorithm of \cite{parsonson2022traffic} works by sequentially iterating through the set of flows and, for each flow, conducting two passes through the candidate source-destination pairs. In the first pass, the packer attempts to match the target node distribution by looping through all pairs, sorted in descending order of the total size of flow information previously assigned, to find a pair which has not yet met its target information load given the target node distribution and total flow arrival duration provided. Failing to find such a node pair, the packer moves to the second pass, whereby it again loops through each sorted pair but now in search of a source-destination combination which, if allocated the flow in question, would not exceed either the source's or the destination's maximum load capacity given any prior flow allocations.

\begin{wrapfigure}{R}{0.5\textwidth}
\begin{minipage}{0.5\columnwidth}
\vspace{-0.8cm}
\begin{algorithm}[H]
\scriptsize
\caption{Vectorised packing algorithm pseudocode.}
\label{alg:vectorised_packer_pseudocode}
\begin{algorithmic}

\STATE \textbf{Input:} $F$, $P$, $\bold{b}^{p, I}_{target}$

\STATE \textbf{Output:} $\bold{b}^{p, I}_{actual}$

\STATE \textbf{Initialise:} $\bold{b}^{p, I}_{actual} = 0 (|P|)$, $\bold{b}^{p, c} = \frac{\text{node capacity}}{2}(|P|)$

\FOR{$f$ in $F$}

    \STATE $\bold{b}^{p, m} = $\texttt{where}$(\bold{b}^{p, c} - f^{s} < 0, 0, 1)$ \texttt{// Generate boolean mask}
    
    \STATE $\bold{b}^{p, I, m}_{target} = \bold{b}^{p, I}_{target}[\bold{b}^{p, m}], \bold{b}^{p, I, m}_{actual} = \bold{b}^{p, I}_{actual}[\bold{b}^{p, m}]$ \texttt{// Mask invalid pairs}
    
    \STATE $\bold{p}^{max} = $ \texttt{argmax}$(2 \cdot \bold{b}^{p, I, m}_{target} - \bold{b}^{p. I, m}_{actual})$ \texttt{// Get furthest pairs}
    
    \STATE $p^{chosen} = $ \texttt{random\_choice}$(\bold{p}^{max})$ \texttt{// Randomly choose pair}
    
    \STATE \texttt{update\_trackers}$(f, p^{chosen})$ \texttt{// Assign flow to pair}

\ENDFOR

\end{algorithmic}
\end{algorithm}
\end{minipage}
\end{wrapfigure}

\secInLine{Vectorised packing algorithm}
We negate the need for separate first and second passes and for nested pair \texttt{for} loops by using vector array operations. We begin by initialising the per-pair remaining capacity vector as the maximum port capacity (half the per-node capacity, since it is split between the source and destination ports) $\bold{b}^{p, c}$. We then sequentially iterate through $f \in F$ and, for each flow $f$, we generate a boolean vector pairs mask $\bold{b}^{p, m}$ which masks out any pair indices $i \in [0, ..., |P|]$ which would exceed their load capacity were they to be allocated the flow in question:

\begin{equation}
    \bold{b}_{i}^{p, m} = 
    \begin{cases}
        0& \text{if } \bold{b}_{i}^{p, c} - f^{s} < 0 \\
        1,              & \text{otherwise}
    \end{cases}
\end{equation}

We then apply this pairs mask to filter out any invalid pairs, thus ensuring that any pair chosen from here on would meet the requirements of the second pass of \cite{parsonson2022traffic} and also reducing the time complexity of the \texttt{argmax} operation below in Equation \ref{eq:argmax_candidate_flows} (since the number of candidate pairs is now reduced). Next, we take the masked candidate pairs' current distances from the target information loads, $\bold{b}^{p,I,m}_{target} - \bold{b}^{p,I,m}_{actual}$, shift them by $\bold{b}^{p,I,m}_{target}$ in order to retain any skewness in $\mathbb{P}^{N}$ for as long as possible given the overall \ac{dc} load specified, and find the pairs in this masked subset which are furthest from their target information loads, $\bold{p}^{max}$:

\begin{equation}
    \label{eq:argmax_candidate_flows}
    \bold{p}^{max} = \text{argmax}\bigg(2 \cdot \bold{b}^{p,I,m}_{target} - \bold{b}^{p,I,m}_{actual}\bigg)
\end{equation}

In order to avoid any bias towards smaller pair indices and create the fade phenomenon in the resultant traffic heat map \cite{parsonson2022traffic}, we randomly choose a pair $p_{chosen} \in \bold{p}^{max}$ to which to allocate the flow $f$, thus meeting the requirements of the first pass of \cite{parsonson2022traffic}. Finally, we update the current total information vector's element for the chosen pair, $\bold{b}^{p_{chosen},I}_{actual}$, and the remaining capacity vector elements $\bold{b}^{p, c}$ for any pairs $p \in P$ which share either a source or a destination with $p_{chosen}$. The pseudocode for this vectorised packer is summarised in Algorithm \ref{alg:vectorised_packer_pseudocode}.

\section{Simulation Setup}



\begin{figure}[!tp]
    \centering
    \includegraphics[width=0.98\textwidth]{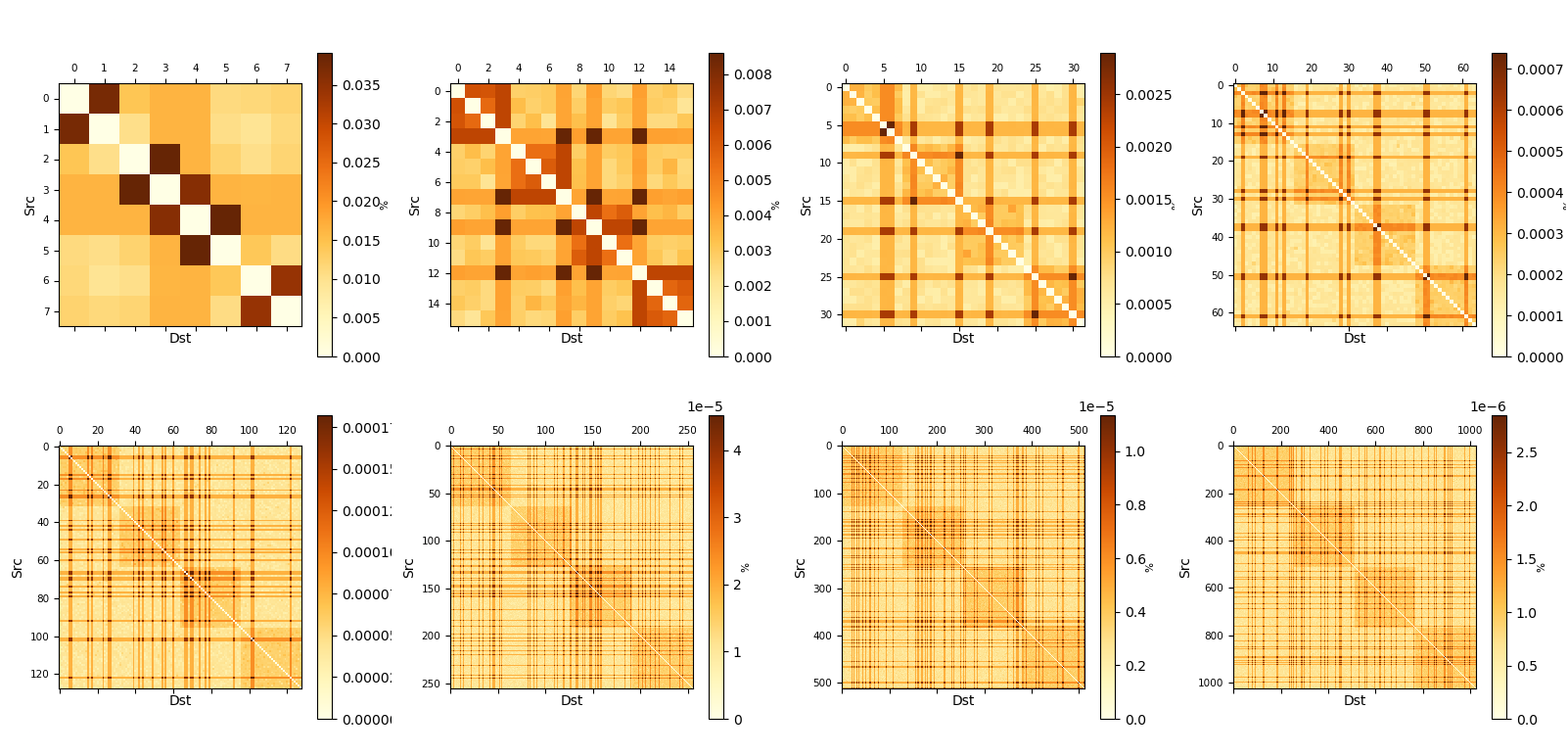}
    \caption{Custom traffic matrix distributions generated with \num{8}, \num{16}, \num{32}, \num{64}, \num{128}, \num{256}, \num{512}, and \num{1024} nodes, where the colour of each source-destination pair corresponds to the fraction of the overall network load it requests.}
    \label{fig:node_dists}
\end{figure}

TrafPy enables the production of custom traffic distributions through the control of a handful of parameters. These include the flow size and inter-arrival time distribution parameters, the node distribution's inter- vs. intra-rack and skew node fractions, and the overall network load. To measure the packing times for the original and vectorised packing algorithms, we generated an assortment of custom traffic patterns typical for a `university' \ac{dc}\footnote{University \acp{dc} service applications such as database backups, distributed file system hosting, and multicast video streaming, with $\approx70\%$ of traffic being inter-rack and $\approx20\%$ of nodes requesting $\approx55\%$ of the traffic load.} as detailed by \cite{parsonson2022traffic} for networks with \num{4} racks and $|N| = \{ 8, 16, 32, 64, 128, 256, 512, 1024 \}$ nodes
(see Figure \ref{fig:node_dists}). We assumed an optical \ac{dc} network with 
with an overall network load rate of $50\%$. For each traffic matrix, we generated $|F|=5 \cdot |N|^{2}$ flows to ensure non-sparse packing. Each packing algorithm was ran on a shared cluster with an Intel Xeon ES-2660 CPU across \num{4} seeds to ensure reliable packing times given the variance in use of the shared cluster, with the 95\% confidence interval bands plotted for any metrics recorded.

\section{Results \& Discussion}





Figure \ref{fig:packing_time_and_jsd}b shows the packing times taken by the original and the vectorised packers when generating the distributions shown in Figure \ref{fig:node_dists}. The vectorised packer achieved a $\approx38\times$ speed-up over the original packer on the $|N|=264$ traffic matrix and $\approx70\times$ on the $|N|=$\num{1024} matrix. 
Although the vectorised algorithm was slightly slower than the original packer on the smallest $|N|=8$ network due to performing a \texttt{where} vector operation on all pairs, the absolute generation time was still $O(s)$ and this additional overhead quickly becomes negligible across $|N|>8$ networks. 

To verify that our proposed vectorised packing algorithm was generating the same node distributions as the original packer used by \cite{parsonson2022traffic}, we measured the \ac{jsd} between the target and the generated node distributions for each algorithm (see Figure \ref{fig:packing_time_and_jsd}c). As expected, both packers deterministically reach the same solution, but the Jensen-Shannon distance will not be exactly $0$ for either due to the incompatability between the $50\%$ network load and the skewed target distribution (see \cite{parsonson2022traffic} for more details).

\textbf{In conclusion}, we have proposed a flow source-destination pair assignment algorithm which makes novel use of vector array operations to achieve orders of magnitude faster traffic generation times than the original algorithm used by \cite{parsonson2022traffic} when generating custom traffic matrices. This work significantly improves the utility of an open source traffic generation framework in order to aid the production of high-fidelity traffic patterns and to test and develop network systems at scale. 

\vspace{0.1cm}

\noindent \scriptsize{\textbf{Acknowledgements}: EPSRC Distributed Quantum Computing and Applications EP/W032643/1; the Innovate UK Project on Quantum Data Centres and the Future 10004793; OptoCloud EP/T026081/1; TRANSNET EP/R035342/1; the Engineering and Physical Sciences Research Council EP/R041792/1 and EP/L015455/1; the Alan Turing Institute; and Horizon Europe Dynamos.}




\bibliographystyle{opticajnl}

\bibliography{sample}

\end{document}